# Knowledge Ecologies in International Affairs: A New Paradigm for Dialog and Collaboration

*"Knowledge is information that changes something or somebody – either by becoming grounds for action or by making an individual (or an institution) capable of different or more effective action."*
Peter Drucker, *The New Realities*

## Introduction

To have command over increasingly complicated social, political, economic and environmental challenges, fragmentary knowledge, or rather the simple accumulation of basic research is inadequate (Kim). International affairs professionals operating in government, academia and the private sector are progressively more aware that access to, and the blending of, interdisciplinary policy-related knowledge is critical to effective problem solving and decision-making. But how can one do so effectively?

This paper examines the concept of knowledge ecologies as a means of addressing this challenge. Although still in their infancy, knowledge ecologies (or knowledge ecosystems) are enabling innovative research, learning and policy solutions. While their intellectual origins can be traced back to the 1990s, it is only in the last few years that we have witnessed the emergence of viable ecologies dedicated to supporting the research and policy process. This pace is certain to increase as the international affairs community embraces new information and communications technologies and reappraises its approach to knowledge creation, management and exchange.

The study of knowledge ecologies covers a broad range of disciplines. In keeping with our subject, so does our paper. We readily acknowledge that in sketching our ideas we do not do justice to any of the disciplines we borrow from. Clearly, more work of both a qualitative and quantitative nature is needed to fully appreciate what impact knowledge ecologies will have on the study of international affairs and the conduct of its community.

In the pages that follow we explore a series of definitions, examine the elements shaping today's knowledge ecologies, and briefly introduce a number of online examples for the reader's reference. Our intention here is not to suggest that knowledge ecologies are a panacea to the real world challenges we face. Instead, we hope to provoke a wider discussion among our peers as to their value and utility.

## Defining Our Subject

The study of knowledge ecologies has grown out of several intellectual and academic disciplines. These have been summarized by Pór (2001) and include complexity science, epistemology, systems thinking, organizational learning, evolutionary science, cognitive science and knowledge management. The study of knowledge ecologies also owes a debt to concepts such as memetics and the noosphere.

The term lends itself to a variety of definitions. Let us begin with a scientific approach. The term "knowledge" dates from the 14th century Middle English *knowlechen*, or acknowledge, and today is used to describe the fact or condition of knowing something with familiarity gained through experience or association, acquaintance with or understanding of a science, art, or technique and no less than the body of truth, information, and principles acquired by humankind. The term ecology is derived from the Greek words *οίκος* (ecos), which means household or institution, and *λόγος* (logos), which means dialog or discourse. By extension, knowledge ecologies can be seen as *a discourse that takes place within a physical or virtual institution*. And just as an ecology is made up of a diversity of inter-connected organisms, minerals and processes that evolve according to their environment, so too does a knowledge ecology consist of a diversity of interdependent and interconnected technologies, processes, entities, strategies, tools, methodologies and communities that adapt to changing circumstances (Young, 2007). Inevitably, the greater the diversity of knowledge, the greater the ecology's adaptability and its resilience to external shock.

From the perspective of an information scientist, a knowledge ecology can also be seen as a dynamic alternative to a traditional ontology, wherein knowledge is created and recreated in multiple contexts and at various points in time (Malhotra, 1999).

To the business leader the term is used to describe a community of practice that generates knowledge using collaborative applications and a bottom-up approach. (Magnan et al, 2007). Such ecologies can be seen in the "skunk works" and "innovation cells" that exist in major industrial concerns, where teams of people are given freedom to work without the fetters of bureaucracy. Until recently the tools they used to interact have been domain specific, and sometimes secret, restricting participants to closed environments of collaboration and knowledge exchange. Increasingly, however, software developers and information architects are pioneering ways to connect disparate communities and the ideas they share, thus enriching these ecologies as well as the innovative and profit-making capabilities of the organization further.

Finally, to an IR scholar, as to anyone active in a wide-ranging field, knowledge ecologies consist of the individuals, institutions and ideas that contribute to the production, collection, analysis, disputation, management, distribution and consumption of research or policy-relevant knowledge. Indeed, the international affairs community is itself a knowledge ecosystem, relying as it does on distinctive academic disciplines for inspiration, ideas and action and, increasingly, on information technologies to develop the disciplines.

Why do these definitions (or perspectives) matter? We would argue that their variety captures the increasingly idiosyncratic nature by which knowledge is generated and shared. Across the board, traditional models of knowledge creation and exchange are being undone. Peer-reviewed journals may contain the best research librarians' money can buy, but their impact is proportional to the audience they receive, and these audiences are likely to dwindle. On the institutional level, the "knowledge capital" of an organization is determined not only by the quality of its research, but also by its willingness to nurture the knowledge commons. Further, a recent report titled *University Publishing in a Digital Age* (Brown et al. 2007) argues that the trend towards open access publishing will nurture new knowledge ecologies, foster new research environments and overhaul the means by which we create and consume educational resources.

The forces driving such processes are interrelated and irresistible. Indeed there is a vast literature – no longer exclusively science fiction – that argues the creation and adoption of information technologies is an evolutionary step for humanity. Information and our attempts to control it are defining life itself (Beniger, 1986). Just as we now accept the web to be a symbiotic part of our natural existence, so too are we apt to accept newer knowledge ecologies as a matter of course.

With this in mind, it is critical to understand the properties of knowledge ecologies and the forces shaping their emergence. By doing so, one can anticipate how these might shape the study of international affairs.

**Dissecting the Ecology**

Broadly speaking, these properties can be examined on both the macro and micro level. On the micro level they consist of the individual nodes that engage in the business of knowledge creation and exchange. Closer analysis here reveals the interests and motivations of these nodes, which can be personal or professional in scope and driven by any number of perspectives from selflessness to self-actualization.

The macro level, however, is concerned with those elements consistent to knowledge ecologies in general. These include, *inter alia*, their culture, organization and design; the technology they employ; their system of collaboration and communication; and the language they employ to achieve their ends. To understand how a knowledge ecology functions from an international studies perspective, it is necessary to consider these elements in greater detail.

*Technology* – technology is the primary driver behind the creation of dynamic web-based ecosystems. The explosion of user-generated content, coupled with the dialectic process of data synthesis and atomization is allowing us to identify patterns and relationships between different disciplines and ideas, which in turn foster new cognitive and semantic approaches to long-standing global problems. Elsewhere, the use of common standards is enabling greater interoperability and information sharing within and between individuals and institutions, as well as improved manipulation, searching, browsing, storage and visualization. Data that was once the preserve of different academic disciplines (e.g. geospatial indices of environmental decay) can now be integrated into a basic text document as a means of underlining policy recommendations, overlaid over maps, merged into ever more revealing mash-ups with multiple and various data sets, and perused and simultaneously modified by parties the world over. Knowledge ecosystems need not invest in expensive technology in order to develop community. In fact, studies show that some of the least costly technologies have proven to be the most fruitful; investments in outreach to potential community members is more critical. (Worthen, 2008). Twitter is but one (albeit extraordinary) example of a low-cost, highly integrated ecology. While some might argue that Twitter, at 140 characters a tweet (or feed), is not expressly in the business of creating new knowledge, a counterpoint may be made that the Library of Congress's acquisition of every Twitter feed since 2006 represents an attempt to capture the widest collection of interconnected sentiments, musings and ideas ever in human history.

Technology is also empowering the architectures of participation, communication and collaboration. Though there are persistent tensions from social networking giants, particularly from Facebook and its ever-changing privacy settings, the overall trend here is toward ever-greater openness, the purpose of which is to generate not only convergent and consensus-oriented solutions, but also diverse interpretations of information based on previously unpredicted contexts and unforeseen assumptions (Malhotra, 1999). Indeed, many technologists argue that the closed architectures being built today stem from previous *technical limitations* and therefore are not to be considered part of the natural way of the world.

*Language and Understanding* – The languages we use to address global challenges are also evolving, albeit at different paces. English is universally acknowledged as the *lingua franca* of international affairs, and for good reason. A shared language augments shared cognition and the search for common solutions to problems such as climate change and weapons proliferation. Moreover, its growth is unfettered by a national academy dedicated to its preservation. As knowledge of the English language evolves, so too will the impetus to communicate in it. This will compel even native English speakers to accommodate new terms and definitions so that they might engage and grow the knowledge ecosystems to which they belong. The essential democratic elements of the English language, as well as its relative simplicity and substantial speaking population also add strength to its position as the language of the Web.

*Societal Context* – Knowledge is invariably a social construct. As ideas jump from one community to another, and from one country or region to the next, they must adapt to prevailing social circumstances if they are to have any currency. It's all very well preaching energy conversation to several million affluent Europeans. How should one do so to a billion plus South Asians, many of whom receive only a couple of hours electricity a day? In a similar vein, democracy may be the best form of government we have; it may also be the one form of government people everywhere aspire to. Nonetheless, recent events have demonstrated at terrible cost that it cannot be transplanted wholesale.

*"Biological" Design* – All ecologies adhere to certain biological principles. Most are self-generating and self-organizing; they form of their own volition and evolve according to internal and external pressures. Second, they operate as networks of relationships. These networks are sustained by communication and the exchange of information, knowledge and ideas. This exchange furnishes the network's meaning and purpose that, in turn, allows the network to define its boundaries. Invariably, these boundaries are permeable. In order to sustain themselves and meet their purpose, networks must remain open to the ideas and influence of others. The intellectual vitality of a knowledge ecosystem serves as a reliable indicator of its future performance, as well as its potential to meet rapidly moving strategic challenges and opportunities (Pór, 2001). In a successful biological ecosystem, genes mutate, organisms are selected, and populations evolve – and in a successful economy, business plans are generated, businesses evolves, and a global economy emerges (moreover, failure to share information and market opacity can lead to economic crashes) (Bray et al, 2008.)

*Organizational and Cultural Drivers* - Knowledge ecologies are a reflection of emerging organizational structures in both the public and private sector. These structures are engineered to ensure maximum resilience and flexibility in environments of high turbulence and uncertainty, as well as to encourage new attitudes and behaviors, especially with regard to knowledge creation and exchange. Thus, to address complex global challenges, the IR community is being compelled to establish analytic nodes across the physical and digital globe. Virtual think tanks will move from being the exception to the norm. As part of the process of adaptation and survival, the IR community will take on a growing range of organizational functions, from policy advocacy to intelligence gathering and analysis (one could go so far as to argue that web-based knowledge ecologies will evolve into the intelligence enterprises of the future). This will require a cultural framework that is as open and flexible as its structure. Academic stovepipes will have little real-world utility. It also requires incentives to encourage greater engagement in the creation and development of new ecosystems.

Existing incentives, especially in academia, appear to be inadequate. A recent study in *Nature* (2006) noted that feedback and peer review tools do not do much to encourage the active sharing of knowledge. In fact, the study notes "most comments were not technically substantive. Feedback suggests that there is a marked reluctance among researchers to offer open comments." (Nature, 2006). In a knowledge ecology, comments and ideas are more likely to be scrutinized constructively. This is because relations between the various analytic nodes are not based on power and authority but rather on a collective sense of curiosity, engagement, and survival based on co-dependence.

*The Virtues of Openness* – Further to the above, the present decade can be called the "open decade" (open source, open systems, open standards, open archives, open everything) just as the 1990s were called the "electronic" decade (e-text, e-learning, e-commerce, e-governance) (Mertu, 2004). But as Peters (2008) argues, the decade of openness has been accompanied by a change of philosophy and ethos that has transformed the marketplace of ideas and the modes of production, collaboration and participation. The drive towards ever-increasing transparency and openness is essential not just to scientific inquiry and progress, but also to the healthy functioning of democracy. However, as Nielsen (2008) points out, to create an open scientific culture that embraces new online tools, two challenging tasks must be achieved: (1) build superb online tools; and (2) cause the cultural changes necessary for those tools to be accepted.

This culture of openness is remaking traditional markets and domains. According to Pór some governments appear to be waking up to the fact that web-enabled technologies make the function of "information gate-keeping" obsolete (Pór, 2001). However, recent work by the OpenNet Initative has shown that governments, particularly in Asia, have expanded their mandate to filter sensitive content both technically and through 'soft controls' such as legal regulation and delegated liability (Deibert, 2008). The basic tension between control and flow remains, even if the most technically savvy are able to procure open information on the web that governments have attempted to filter.

In an example of the growth of openness that relates to the IR community, the growing popularity of open source intelligence appears to be evidence of both a cultural and organizational shift that values knowledge as an imperative to action rather than occlusion. We are witnessing the emergence of an intellectual commons, one that is sustained by the efforts of researchers, writers, academics, bloggers and encyclopedists, as well as a global telecommunications network.

Knowledge is increasingly seen as a public good. Similarly, those who nurture the commons by enabling the sharing of ideas are increasingly seen as *public servants.* Their role will only grow in importance as we begin to tackle problems of a global, systemic nature. While their motivations may differ, as we have seen with Wikipedia and other online communities, most will be committed volunteers. As Benkler (2007) reminds us:

> "Together, these three characteristics [ubiquitous computing, knowledge as a public good, modularity of technical architectures and social dynamics of knowledge production] suggest that the patterns of social production of information that we are observing in the digitally networked environment are not a fad. They are, rather, a sustainable pattern of human production given the characteristics of the networked information economy."

Without access to a global knowledge commons governments and individuals can only address problems from necessarily narrow (and sometimes selfish) perspectives rather than through a process of collective understanding and a shared responsibility for the outcomes. The networks that furnish the commons are moving towards ever-greater openness and transparency. Only by doing so can the network's purpose be communicated and its knowledge ecology nurtured and sustained.

*The Convergence of Work and Learning* – It seems almost trite to acknowledge that a Masters or PhD is no guarantee of lifetime employment. And yet, thousands of IR students graduate each year with little or no knowledge of the 21st century literacies vital to their places of work. For those in the business of training and hiring, the skills shortage is palpable. Efforts like the American Library Association's Information Literacy Competency Standards for Higher Education (2000) go far, but only if educators are willing to assign value to this type of competency.

Knowledge ecologies are alerting participants not only to the theoretical trends shaping their disciplines, but also to the practical skills they must acquire in order to succeed. Looked at differently, knowledge ecologies operate as biological sensors, alerting their constituents to the properties they must adopt to ensure their relevance and survival. Leveraging knowledge and experience into a process of continuous learning and thoughtful reflection moves us toward insight. And insight is one of the ingredients of true innovation (Mamprin, 2006), whether in research or policy-making.

*Changing Work Patterns* – Changing work patterns are a consequence of the organizational and cultural changes outlined above. A virtual think tank can work on a perpetual cycle of data collection and analysis. Field research need not be packaged into edited volumes (whose utility to the international studies community may be negligible by the time they reach the library's bookshelf). Similarly, insights gleaned from the morning's headlines need not wait for a conference paper or presentation. Assumptions, hypotheses, opinions and facts can be uploaded to the web and scrutinized in near real time by individuals populating the same knowledge ecologies.

*The Culture of Knowing* – Knowledge ecologies are emerging thanks to our individual and collective desire to know. Their advocates are noted for the passion with which they consume and discuss ideas from disparate fields. Indeed, the local and global contexts within which knowledge ecologies evolve necessitate greater intelligence, awareness and understanding on the part of its members. Thus, knowledge ecologies are expeditionary in nature; they are inspired by the prospect of discovery and shaped through exploration (Mason et al., 2003). As they become smarter, so too do they become more capable and effective.

Just as the world of work will not forgive a graduate lacking in practical skills, so too will the culture of knowing refuse to forgive a willing ignorance of scientific advances or sociological trends. Knowledge ecologies compel us not only to know more, but also to *know knowledge* and the forces that enable its creation and exchange. Introductory epistemology would not be out of place on a progressive IR syllabus eager to encourage younger scholars to know the world through multiple perspectives. Indeed, it is imperative that we understand how concepts and ideas are interpreted in different settings. Knowledge ecologies teach us that global challenges can be understood in different contexts at the same time. Thus, when speaking of climate change, one must also ask *whose climate*? And *what change*? In a knowledge ecology, the basis for cooperation and survival is differentiation *and* similarity between the individuals and perspectives that exist within it (Malhotra, 1999).

*Shifting Lines of Power and Authority* – Today's knowledge dynamics will continue to undermine traditional power structures. Knowledge ecologies demonstrate how ideas, insights and opinions flow to where they best needed or appreciated. No doubt, this will shape the loyalties of those who have knowledge to contribute. The maverick policymaker may find more use for his or her ideas in an ecology of strangers than in the stilted environment they work in. Thus, knowledge ecologies will become marketplaces for ideas and cohesive mechanisms of change for institutions and professions.

**Ecologies as the Evolution of Knowledge Management**

Knowledge ecologies are both a compliment to, and the successor of, traditional knowledge management (KM) approaches. KM remains a powerful enabler of organizational efficiency and effectiveness. Its value is all too apparent to those organizations that have succeeded in implementing effective KM practices. Regrettably, most have failed to do so, assuming that solutions lie in the technologies they adopt rather than the people they employ. Much remains to be done here. However, no discipline stands still for too long. Knowledge ecologies are slowly supplanting knowledge management as the principle by which the international studies community will increasingly marshal its knowledge assets. Why?

In an era of discontinuous change, it is necessary to move beyond technological frameworks designed around predictive rules of engagement to complex adaptive systems engineered to anticipate surprise (Malhotra, 1999; Bray et all, 2008). Put differently, knowledge ecologies move towards *adaptation* while knowledge management systems tend toward *optimization*. The former enables responsiveness; the latter inevitably spells redundancy. And whereas knowledge management seeks to harness know-how and know-what, knowledge ecologies provide the context and enables trust and collaboration.

Table 1 below summarizes the differences between the two approaches:

|  | Knowledge Management | Knowledge Ecologies |
|---|---|---|
| Provision | Provides actionable information and opportunity typically gathered from a finite network<br><br>Provides explicit knowledge in the form of guidelines, best practices, etc. | Provides the context, synergy and trust necessary to generate information and lessons learned, recognize opportunity and turn them into knowledge and action; sources are potentially infinite<br><br>Enables the sharing of tacit knowledge through dialog and collaboration |
| Structure | Emphasis on *hard* architectures that guarantee intellectual asset protection, fixed knowledge objects and standardized rules of knowledge exchange that can be audited and improved | Emphasis is on *organic* structures conducive to pattern recognition, sensemaking, prototyping, adaptation and feedback, and self-generating rules approaches to knowledge creation and use |

| Enablers | Technology; knowledge is embedded in databases and knowledge creation occurs through *access* | Networks of individuals and institutions; knowledge is embedded in people and knowledge creation occurs through *interaction* |
|---|---|---|
| Orientation | Bottom-line oriented – it allows us to see the challenges and opportunities for assessing, organizing, portraying and profiting from knowledge | Community oriented – it allows us to recognize the effort needed to grow and sustain networks of relationships from which knowledge can emerge |
| Focus | Dedicated to *formulating policies* on knowledge distribution and access, and ways to ensure compliance with the | Dedicated to an ongoing *dialog on policy* to enable shared cognition and understanding; knowledge ecologies seek alignment but do not insist on control |
| Physical Properties | Emphasis on intellectual matter, and thus knowledge *particles* in the form of rules, best practices, documents, FAQs, etc. | Emphasis on intellectual energy, thus knowledge waves or relationship. These are facilitated through trust, dialog, opinions, innovation, and creativity |

(Adapted from George Pór and the Community Intelligence Labs, 1997.)

**Knowledge Ecologies in Practice**

So much for the theory; what of ecologies in practice? A number of international affairs-related websites can be identified as emerging knowledge ecologies. While they may have started life as static web services, they are beginning to demonstrate many of the macro-elements noted above. These ecologies are emerging as a result of a turbulent international environment. According to Dumaine (2008), this environment is characterized by:

- The emergence of non-state actors as drivers of global security challenges
- The proliferation of alternative information sources and communication tools
- The unpredictability and volatility of world events

- The blurring of foreign and domestic issues
- Greater complexity and interconnectedness in the physical and virtual environments we inhabit
- Asymmetries affecting strategy and conflict
- Organizational disabilities

Managing these challenges requires as yet unseen levels of research, cooperation, coordination, and information sharing at the international, regional and national level (Nielsen, 2008). The basis for much of this will inevitably be realized online. The services we examine below demonstrate some of the macro-characteristics outlined above. As time passes, we suspect they will evolve into fully-fledged ecologies dedicated to leveraging the insights and expertise of their users.

One of the oldest active international affairs portals on the web, Relief Web (www.reliefweb.net) actively encourages the participation of the IR community. The site welcomes contributions from NGOs, UN agencies, governments, think tanks and the media. Drawing on the contributions of others, Relief Web publishes situation reports, appeals, policy documents, analyses, press releases and maps in support of the humanitarian affairs community and their operations worldwide. By encouraging greater information sharing on the part of its users, Relief Web hopes to improve the collective understanding and response capacity of the community as a whole. While Relief Web does a great service collecting and pushing content, it would benefit from greater community involvement and sensemaking tools.

In 2004, The Centre for International Governance Innovation (CIGI) launched IGLOO (www.igloo.org) which operated as a "network of networks". IGLOO enabled the creation of subject and region-specific communities of practice by inviting participants to make use of its community development software. To date, dozens of academic and policy-oriented communities comprising thousands of members have been developed using this tool. IGLOO's strength lay in the motivation of its users and the usability of its software tools. Anyone can launch a network and, depending on the access restrictions they define, open participation to as broad or as narrow a community of participants as they deem necessary. Naturally, the more open the network, the greater the ecology of input and ideas they will sustain. Inevitably, one must engage the networks they are interested in if they are to benefit from, or contribute to their work. This is an inevitable trade off that ensures both user commitment and transparency. In 2008, IGLOO was spun off as a commercial venture (http://www.igloosoftware.com/) with the mandate to overcome information trapped in silos.

The International Relations and Security Network (ISN – www.isn.ethz.ch) is an altogether different ecology. Established in 1994, it operates as an information, IT and educational services provider to the international affairs community. The ISN exists to encourage greater knowledge sharing among international affairs professionals. Given that many of these operate in an institutional framework, its primary participating unit is the organization rather than the individual. Data is carefully tagged and indexed to ensure maximum findability. However, while users are able to contribute to the development of this ecology by adding research of their own, there are few mechanisms for online dialog or collaboration. Users can post comments on news analysis articles but not yet on policy or research papers. The ability to do so would enable the ISN to profit from the insights and experience of their users more directly.

The ISN has also developed communities of its own, many of which are openly accessible to individual users. Here too, however, collaboration is oriented on an institutional rather than an individual level. The more engaged the institutional partner, the likelier it is to profit from its membership in the ISN's communities of practice. The effectiveness of these communities as incubators of innovative research and policy solutions will need to be addressed in time.

Pushing the envelope further, a recent emerging ecology is the Global "Energy and Environment Strategic Ecosystem" (EESE) - http://globaleese.org/. In 2007 the US Department of Energy's Office of Intelligence and Counterintelligence established an Energy and Environmental Security Directorate. In order to understand future problems and consequences more fully, this Directorate initiated the formation of a truly global, interdisciplinary, collaborative and adaptive network to obtain the necessary insights on complex connections and dependencies associated with energy and environmental concerns.

Given the complex scientific and societal issues associated with today's energy and environmental concerns, formulation of an effective policy solution requires a global strategic intelligence capability that employs an ecosystem approach to identifying present and future challenges, as well as potential solutions to these encroaching dilemmas. The Global EESE network was thus designed to increase dialog and openness, surface unanswered questions and amplify weak signals related to energy and environmental concerns. However, despite its initial successes, Global EESE was not able to anticipate political shifts that would undermine its funding. At this point, it is not clear whether it will continue to grow.

Inviting people to "join the dialogue" and "brag a little" the developers corner of Data.gov, the flagship of the Obama Administration's Open Government Initiative, is fostering a knowledge ecology around machine-readable datasets that are generated and held by the Federal Government. By sharing mashups and other visualizations that were not possible just a few years earlier, Data.gov has become a focal point for participatory democracy and creative use of data to further policy goals and societal understanding. Helping to further evolve Data.gov, Ideascale (http://datagov.ideascale.com/) offers a service for developers of all stripes to share and vote on new uses of government data. While initially criticized by some as a data-dumping ground, it seems now that Data.gov is both a catalyst and emerging knowledge ecology.

**Conclusion**

The services outlined above are in various stages of development. However, they all demonstrate the properties of emergent knowledge ecologies and are evolving rapidly to meet the broader needs of the communities they serve. To ensure their continued relevance, they must increase incentives for participation while at the same time enabling a shared sense of community and purpose, despite the geographical distribution of community members, as well as their anonymity to one another.

Chris Pallaris, International Relations and Security Network, ETH Zurich
Sean S. Costigan, Institute of Foreign Policy Studies, University of Calcutta